\documentclass{rmf-d}
\usepackage{nopageno,rmfbib,multicol,times,epsf,amsmath,amssymb,cite}
\usepackage[latin1]{inputenc}
\usepackage[]{caption2}
\usepackage{graphics}
\usepackage{hyperref}
\usepackage{xcolor}
\usepackage[demo]{graphicx}
\usepackage{wrapfig, blindtext}
\usepackage{comment}
\usepackage{tabularx}
\usepackage{float}
\usepackage{capt-of}
\def\rmfcornisa{Research OR Education of Physics \hfill\rmf\ {\bf ??} (*?*) ???--???
\hfill MES? A\~NO?}

\def\rmfcintilla{{\it Rev.\ Mex.\ Fis.\/} {\bf ??} (*?*) (????) ???--???}
\clearpage \rmfcaptionstyle 
\begin{document}
\title{Properties of low-lying charmonia and bottomonia from lattice QCD + QED
\vspace{-6pt}}
\author{J. Koponen}
\address{PRISMA+ Cluster of Excellence \& Institute for Nuclear Physics,\\
Johannes Gutenberg University of Mainz, D-55128 Mainz, Germany}
\author{B. Galloway, D. Hatton, C. T. H. Davies}
\address{SUPA, School of Physics and Astronomy, University of Glasgow, Glasgow, G12 8QQ, UK}
\author{G. P. Lepage}
\address{Laboratory for Elementary-Particle Physics, Cornell University, Ithaca, New York 14853, USA}
\author{A. T. Lytle}
\address{Department of Physics. University of Illinois, Urbana, IL 61801, USA}

\maketitle
\recibido{day month year}{day month year
\vspace{-12pt}}
\begin{abstract}
  \vspace{1em}
  The precision of lattice QCD calculations has been steadily improving for some time and is now approaching,
  or has surpassed, the 1\% level for multiple quantities. At this level QED effects, i.e. the fact that
  quarks carry electric as well as color charge, come into play. In this report we will summarise results from
  the first lattice QCD+QED computations of the properties of ground-state charmonium and bottomonium mesons
  by the HPQCD Collaboration.
  \vspace{1em}
\end{abstract}
\keys{Charmonium, bottomonium, lattice QCD, lattice QCD+QED  \vspace{-4pt}}
\pacs{   \bf{\textit{14.40.Lb, 14.40.Nd, 12.38.Gc}}    \vspace{-4pt}}
\begin{multicols}{2}

\section{Introduction}

Lattice QCD has been the gold standard for calculating properties of hadrons in Standard Model for a
long while \cite{HPQCD:2003rsu}. For many quantities, such as masses and decay constants of ground-state
pseudoscalar mesons, calculations have now reached, or surpassed, statistical precision of 1\%. This
precision of modern lattice QCD results means that sources of small systematic uncertainty that could
appear at the percent level need to be understood and quantified. Here we focus on QED effects.

In the following section we briefly introduce the lattice QCD setup, as well as describe how we include QED in
the calculation. In section \ref{sec:results} we summarise our results on charmonium and bottomonium hyperfine
splittings and decay constants published in \cite{Hatton:2020qhk, Hatton:2020vzp, Hatton:2021dvg}.

\section{Lattice calculation}

We use gluon field configurations generated by the MILC collaboration \cite{MILC:2012znn, Bazavov:2017lyh}.
We use 17 different ensembles: six different lattice spacings from very coarse ($a\approx 0.15\mathrm{~fm}$) to
exafine ($a\approx 0.03\mathrm{~fm}$), and a range of light quark masses (including close to physical masses)
to control the chiral extrapolation. Most ensembles have $2 + 1 + 1$ flavours, i.e. light, strange and charm
quarks in the sea (with degenerate $u$ and $d$ quarks whose mass is $m_l=(m_u+m_d)/2$). However, we use one
ensemble with $n_f = 1 + 1 + 1 + 1$, where both $u$ and $d$ quarks have their respective physical masses.

The Highly Improved Staggered Quark (HISQ) action \cite{Follana:2006rc}, which removes tree-level $a^2$
discretisation errors, is used for both sea and valence quarks. For heavy quarks the `Naik' term is adjusted
to remove $(am)^4$ errors at tree-level, which makes the action very well suited for calculations that involve
$c$ quarks. For the $b$ quarks we use the so called heavy-HISQ method \cite{McNeile:2011ng}, i.e. do the
calculation at several heavy valence quark masses $m_h>m_c$ to extract quantities at the physical $b$ mass.

\subsection{QED on the lattice} 

To study the systematic effects related to the fact that quarks carry both electric and color charge,
we have to include QED in our QCD calculation. We use quenched QED, i.e. we include effects from the
valence quarks having electric charge (the largest QED effect) but neglect effects from the electric
charge of the sea quarks. In short, the calculation goes as follows (see \cite{Hatton:2020qhk} for
details):
\begin{itemize}
\item Generate a random momentum space photon field $A_{\mu}(k)$ for each QCD
  gluon field configuration and set zero modes to zero using the QED$_L$ formulation (QED in finite box).
\item Fourier transform $A_{\mu}$ into position space. The desired $U(1)$ QED field is then the
  exponential of $A_{\mu}$, $\mathrm{exp}(ieQA_{\mu})$, where $Q$ is the quark electric charge in units
  of the proton charge $e$.
\item $c$ and $b$ lattice quark masses have to be tuned separately in pure QCD and QCD+QED so that
  $J/\psi$ and $\Upsilon$ masses match experiment.
\end{itemize}

\subsection{Extraction of energies and decay constants}

We calculate the quark-line connected correlation functions of pseudoscalar and vector mesons on each ensemble
and use a multi-exponential fit to extract amplitudes and energies:
\begin{equation}
  C_{\textrm{2-point}}(t)=\sum_iA_i\Big(\mathrm{e}^{-E_it}+\mathrm{e}^{-E_i(L_t-t)}\Big).
\end{equation}
The decay constants are related to the ground state ($i=0$) amplitude and meson mass:
\begin{equation}
  f_P=2m_q\sqrt{\frac{2A_0^P}{(M_o^P)^3}},\quad
  f_V=Z_V\sqrt{\frac{2A_0^V}{M_0^V}}.
\label{eq:fP_fV}  
\end{equation}
The renormalisation constant $Z_V$ is needed to match the lattice vector current to that in continuum
QCD, as we use a non-conserved lattice vector current \cite{Hatton:2019gha}. The current used for the
decay constant $f_P$ is absolutely normalised, and no renormalisation factor is required.

We then take the results at different lattice spacings and extrapolate to the continuum, taking into account
$(am_q)^{2n}$ and $(a\Lambda)^{2n}$ discretisation effects. Terms that allow for mistuned sea
quark masses are also included. For bottomonium, we map out the dependence in quark mass to extract the
result at the physical $m_b$.

\section{Charmonium and bottomonium}
\label{sec:results}

Let us now summarise our results on charmonium and bottomonium hyperfine splittings and decay constants.

\subsection{Hyperfine splitting}

\begingroup
\begin{center}
\includegraphics[trim={0 0.5cm 0 0},clip,width=0.99\columnwidth]{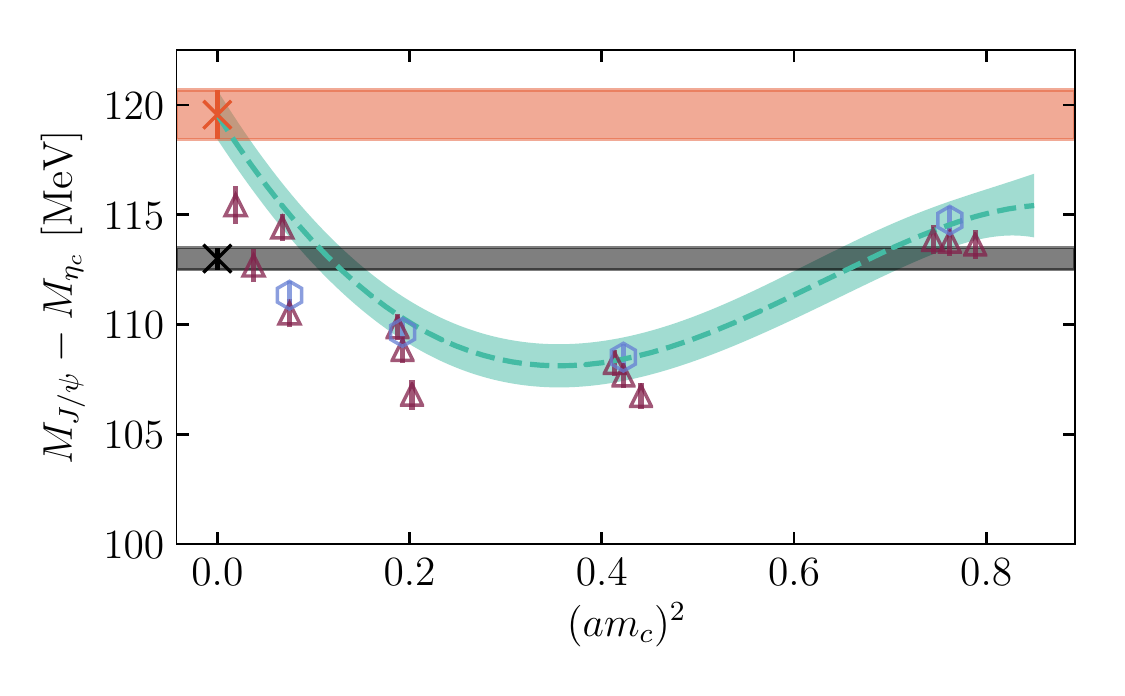}
\captionof{figure}{Charmonium hyperfine splitting as a function of lattice spacing.
  This figure is from \cite{Hatton:2020qhk}.\label{fig:charmonium_hf}}
\end{center}
\endgroup

In figure \ref{fig:charmonium_hf} we plot the hyperfine splitting as a function of lattice spacing, the
blue hexagons and violet triangles showing our results on different ensembles in pure QCD and in QCD+QED
respectively. Our extrapolation to the continuum and to physical quark masses is shown by the turquoise
error band. The red error band gives our physical result, and the black cross and the black error band
show the average experimental result from Particle Data Group \cite{PDG2018}.
Our final QCD+QED result for the charmonium hyperfine splitting is $M_{J/\psi}-M_{\eta_c}=120.3(1.1)\mathrm{~MeV}$.

For the first time we see a significant, 6$\sigma$ difference between the experimental average and
a lattice calculation. Note that quark-line disconnected correlation functions are not included in
the lattice calculation. The difference between our result and the experimental result is then taken
to be the effect of the $\eta_c$ decay to two gluons (prohibited in the lattice calculation):
$\Delta M_{\eta_c}^{\textrm{annihln}}=+7.3(1.2)\mathrm{~MeV}$. 

\begingroup
\begin{center}
\includegraphics[trim={0 0.2cm 0 0.1cm},clip,width=0.99\columnwidth]{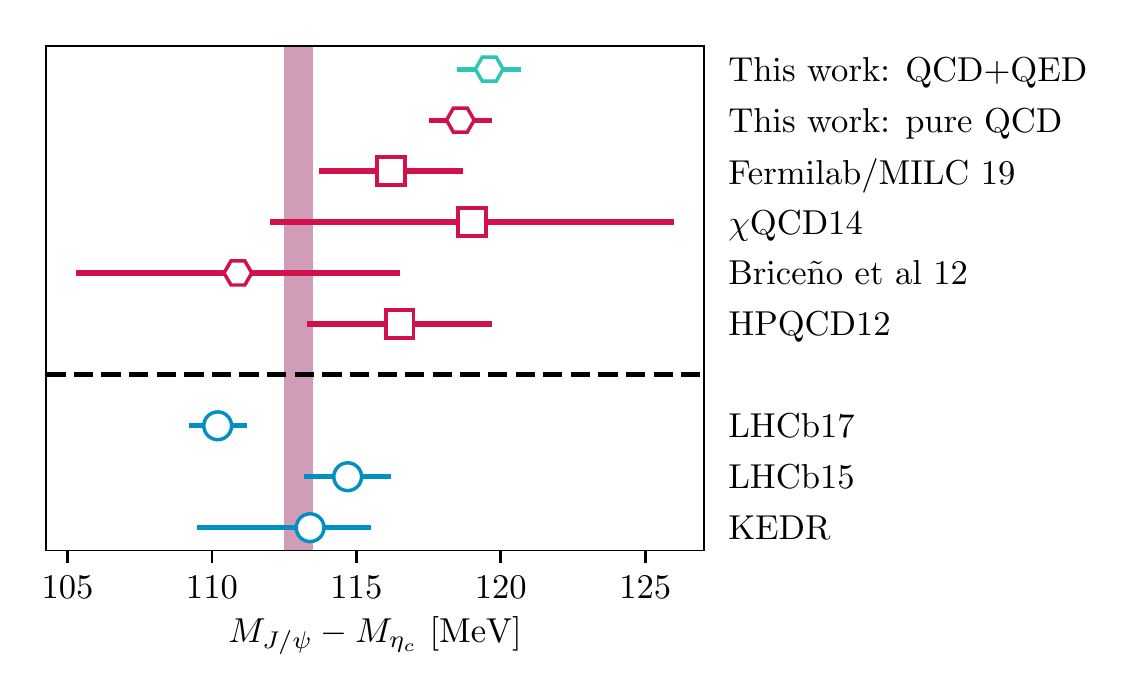}
\captionof{figure}{Charmonium hyperfine splitting. This figure is from \cite{Hatton:2020qhk}.\label{fig:charmonium_hf_comp}}
\end{center}
\endgroup

In figure \ref{fig:charmonium_hf_comp} we compare our result for $M_{J/\psi}-M_{\eta_c}$ with other
lattice QCD results as well as with experimental results that measure this difference. The results
are from the following piblications: Fermilab/MILC \cite{DeTar:2018uko}, $\chi$QCD \cite{Yang:2014sea},
Briceno \cite{Briceno:2012wt}, HPQCD \cite{Donald:2012ga}, LHCb \cite{LHCb:2014oii, LHCb:2016zqv} and
KEDR \cite{Anashin:2014wva}. The PDG average, shown as the purple error band, is obtained from taking
the differences of the PDG $J/\psi$ and $\eta_c$ masses rather than only from experiments that directly
measure the splitting.

\begingroup
\begin{center}
\includegraphics[width=0.9\columnwidth]{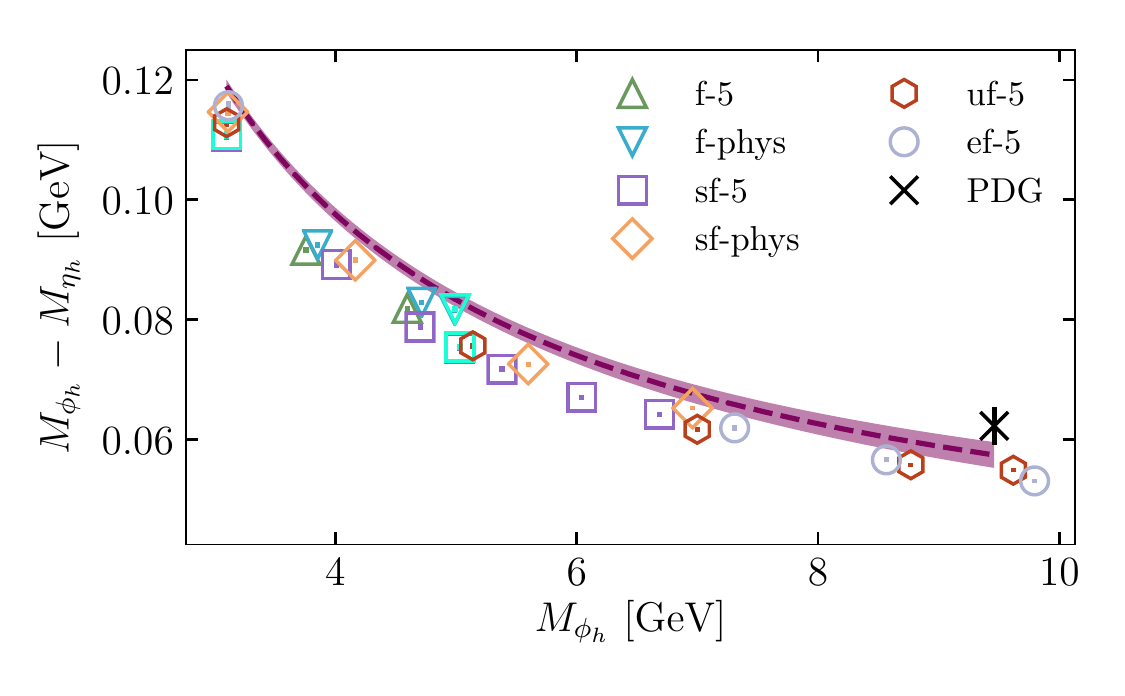}
\captionof{figure}{Bottomonium hyperfine splitting. This figure is from \cite{Hatton:2021dvg}.\label{fig:bottomonium_hf_splinefit_wQED}}
\end{center}
\endgroup

To study the bottomonium hyperfine splitting, we map out the dependence in $m_h$ to extract the result at physical $m_b$.
This is illustrated in figure \ref{fig:bottomonium_hf_splinefit_wQED}, where we plot our results on different lattice
ensembles as a function of the heavy vector meson mass $M_{\phi_h}$ (which is a proxy for the heavy quark mass).
The error band shows the extrapolation to the continuum, and the black cross shows the experimental average from
Particle Data Group \cite{PDG2020}.

Our QCD+QED result for bottomonium hyperfine splitting is $M_{\Upsilon}-M_{\eta_b}=57.5(2.3)(1.0)\mathrm{~MeV}$.
The missing quark-line disconnected contributions (allowed for by the second uncertainty) are
expected to be smaller for bottomonium than charmonium, and here we find good agreement with experiment.

\begingroup
\begin{center}
\includegraphics[width=0.99\columnwidth]{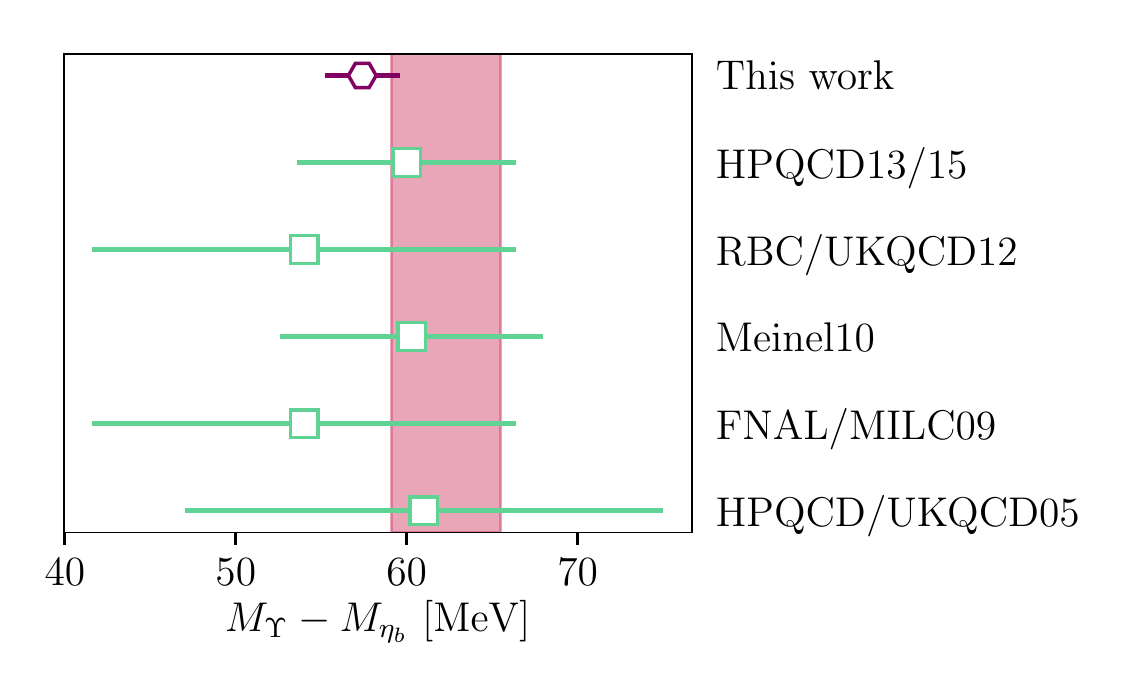}\\
\includegraphics[width=0.99\columnwidth]{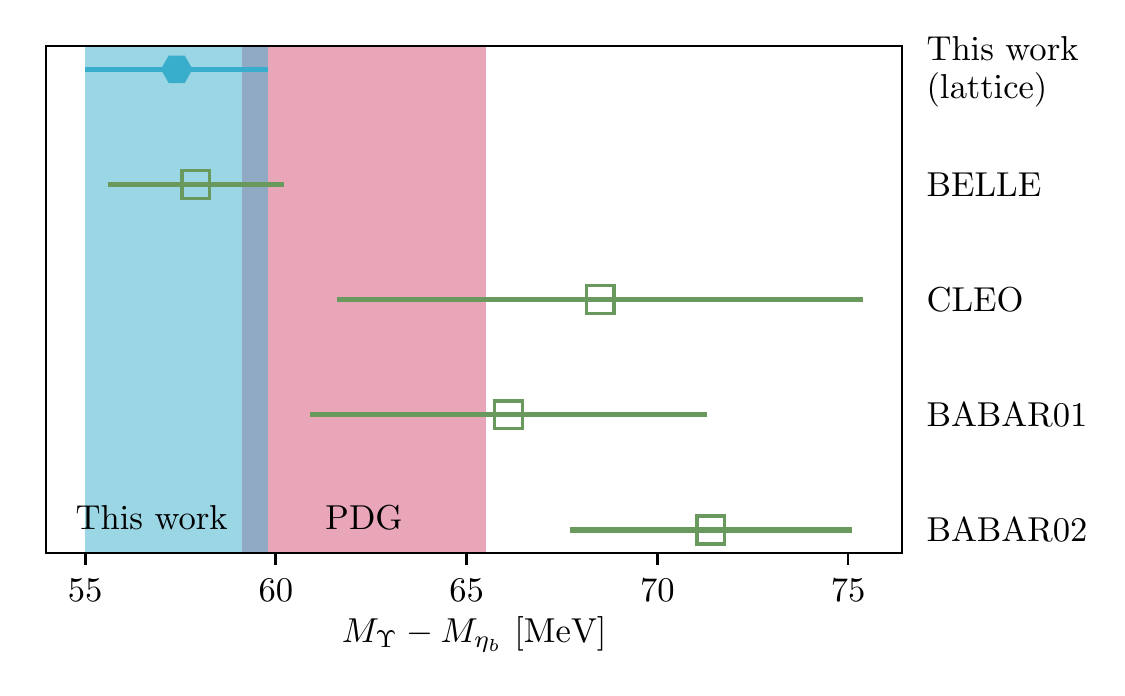}
\captionof{figure}{Bottomonium hyperfine splitting. These figures are from \cite{Hatton:2021dvg}.\label{fig:bottomonium_hf_comparison}}
\end{center}
\endgroup

We compare our results to other lattice QCD results and experimental results in figure \ref{fig:bottomonium_hf_comparison}.
These results are from the following publications: lattice calculations by HPQCD/UKQCD \cite{Gray:2005ur},
Fermilab/MILC \cite{Burch:2009az}, Meinel \cite{Meinel:2010pv}, RBC/UKQCD \cite{RBC:2012pds} and HPQCD \cite{Dowdall:2013jqa}, 
and experimental results from Belle \cite{Belle:2012fkf}, CLEO \cite{CLEO:2009nxu} and BaBar \cite{BaBar:2009xir, BaBar:2008dae}
as well as the experimental average from Particle Data Group \cite{PDG2020}. All lattice calculations show good agreement,
but there is some tension between the different experimental results with our value favouring (but not significantly) the
most recent lower result from Belle.

\subsection{Decay constants}

The decay constant of a pseudoscalar meson $P$ (e.g. $\eta_c$ or $\eta_b$) is
defined in terms of the axial current as
\begin{equation}
\langle 0|A_{\alpha}|P\rangle =p_{\alpha}f_P.
\end{equation}
Using the PCAC relation this can be written as
\begin{equation}
\langle 0 |\bar{\Psi}_q\gamma_5\Psi_q| P\rangle = \frac{(M_0^P)^2}{2m_q}f_P.
\end{equation}
For a vector meson (e.g. $J/\psi$ or $\Upsilon$) the vector decay constant is defined through the vector
current
\begin{equation}
\langle 0|\bar{\Psi}_q\gamma_{\alpha}\Psi_q|V\rangle=f_VM_V\epsilon_{\alpha},
\end{equation}
where $\epsilon$ is the polarisation vector of the meson.

The tensor decay constant of the vector meson is
\begin{equation}
\langle 0|\bar{\Psi}_q\sigma_{\alpha\beta}\Psi_q|V\rangle=if^T_V(\mu)(\epsilon_{\alpha}p_{\beta}-\epsilon_{\beta}p_{\alpha}).
\end{equation}
Note that the tensor decay constant is scale- and scheme-dependent, unlike the vector
decay constant $f_V$.

The decay constants can be written in terms of meson masses and amplitudes --- see Eq. \eqref{eq:fP_fV} along with
\begin{equation}
f_T=Z_T\sqrt{\frac{2A_0^T}{M_0^V}},
\end{equation}
using amplitudes from a tensor-tensor correlation function.

Our results for the charmonium pseudoscalar and vector decay constants $f_{\eta_c}$ and $f_{J/\psi}$ on different lattice
ensembles are plotted as a function of the lattice spacing in figure \ref{fig:charmonium_fP_fV}. The error band shows
our extrapolation to the physical point. For $f_{\eta_c}$, the black cross shows the result from an earlier lattice
calculation by the HPQCD collaboration \cite{Davies:2010ip}, whereas for $f_{J/\psi}$ the black cross shows the result
determined from the experimental average for $\Gamma(J/\psi\to e^+e^-)$. Our QCD+QED results at the physical point are \cite{Hatton:2020qhk}
$f_{J/\psi}=410.4(1.7)\mathrm{~MeV}$, $f_{\eta_c}=398.1(1.0)\mathrm{~MeV}$ and $f_{J/\psi}/f_{\eta_c}=1.0284(19)$.

\begingroup
\begin{center}
\includegraphics[trim=0 5mm 0 4mm, clip, width=0.99\columnwidth]{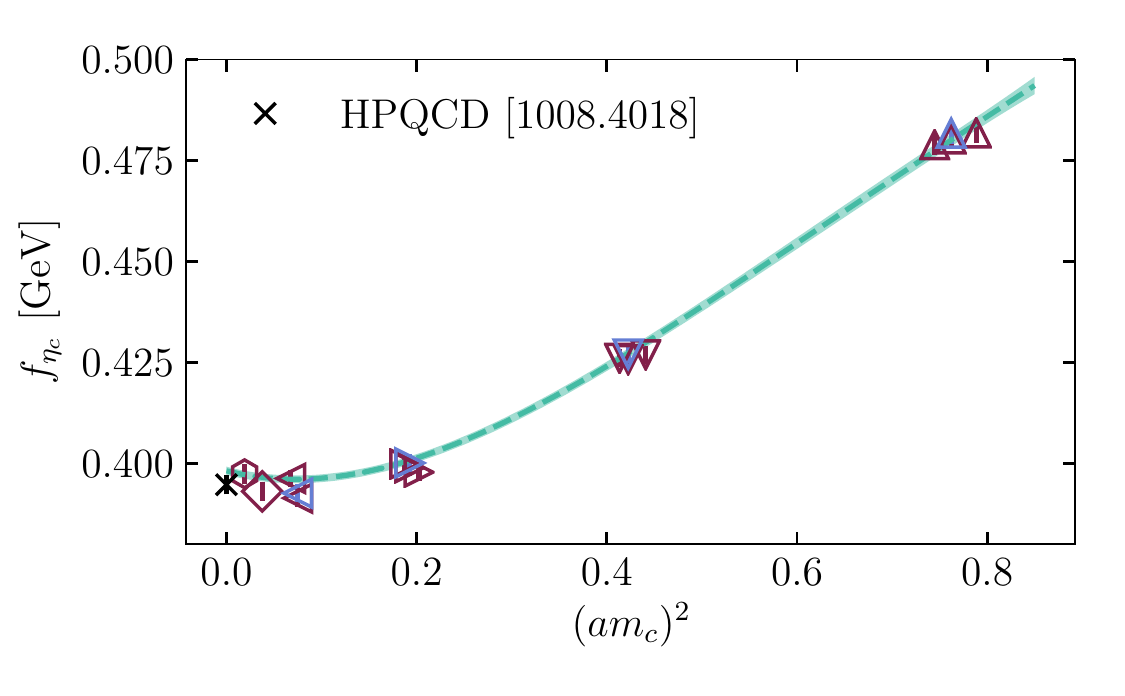}\\
\includegraphics[trim=0 5mm 0 0, clip, width=0.99\columnwidth]{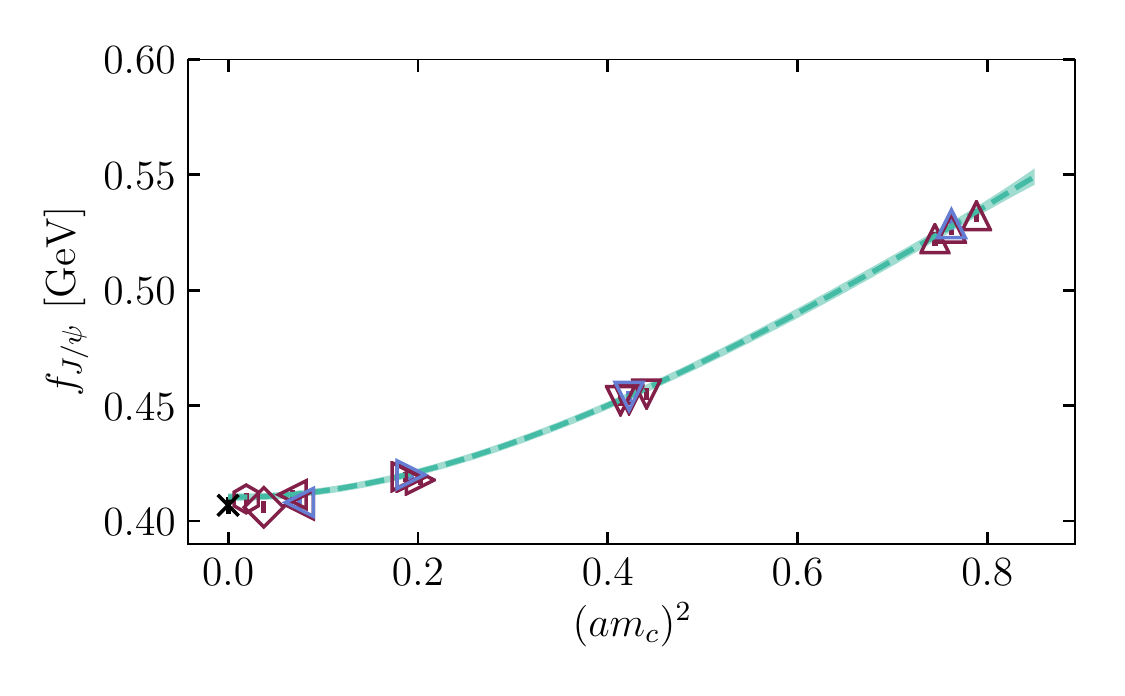}
\captionof{figure}{Charmonium decay constants. These figures are from \cite{Hatton:2020qhk}.\label{fig:charmonium_fP_fV}}
\end{center}
\endgroup

\begingroup
\begin{center}
\includegraphics[width=0.94\columnwidth]{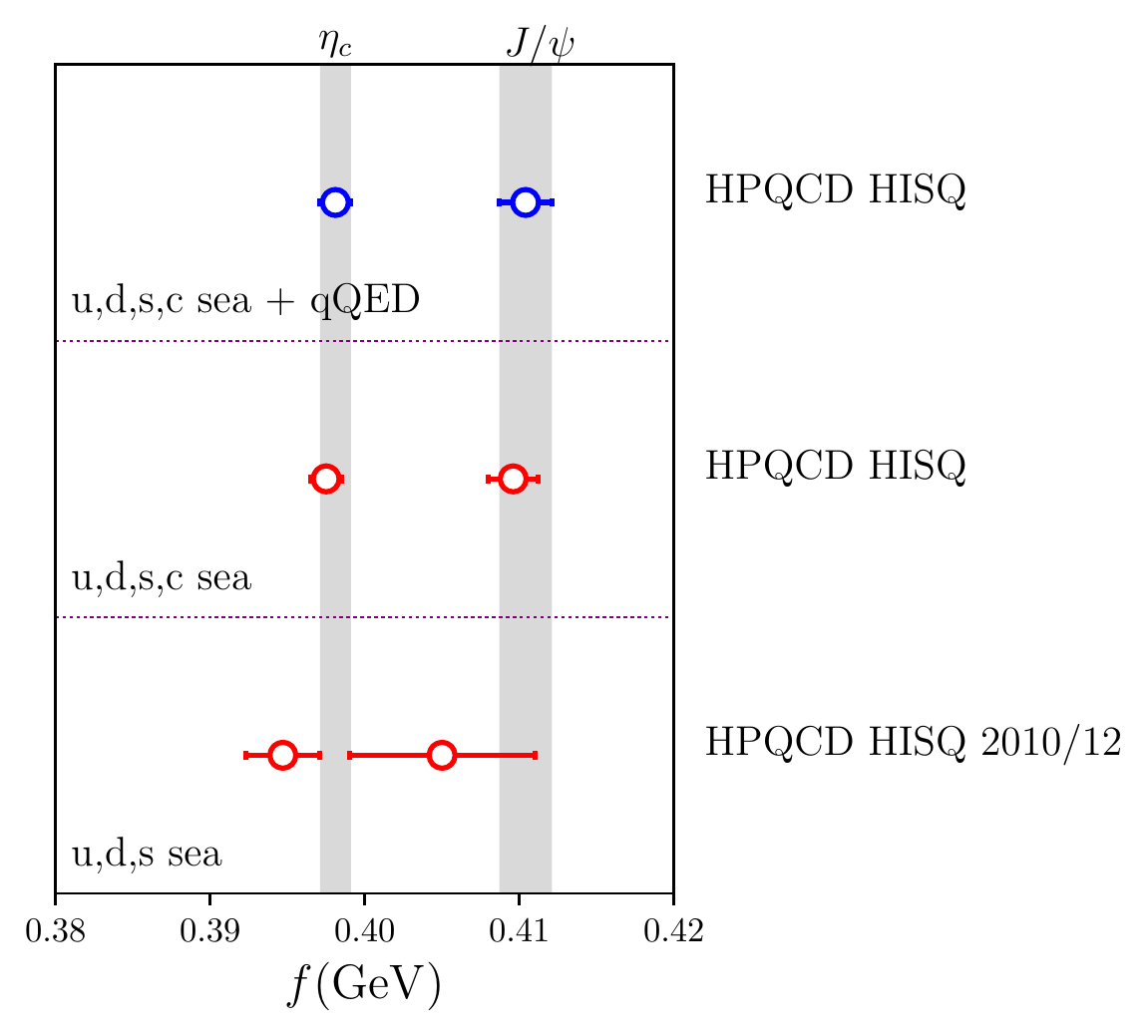}
\captionof{figure}{Charmonium decay constants.\label{fig:charmonium_f_comparison}}
\end{center}
\endgroup

The decay constants from the QCD+QED calculation are compared with the pure QCD results in figure
\ref{fig:charmonium_f_comparison}. The QED effects are very small, but at this precision they have to be taken
into account. Figure \ref{fig:charmonium_f_comparison} also compares these new results to an earlier lattice
calculation by the HPQCD collaboration that had only $u$, $d$ and $s$ quarks in the sea
\cite{Donald:2012ga, Davies:2010ip}. The improvement in the precision highlights how far lattice calculations
have come.

\begingroup
\centering
\includegraphics[trim=0 5mm 0 0, clip, width=0.99\columnwidth]{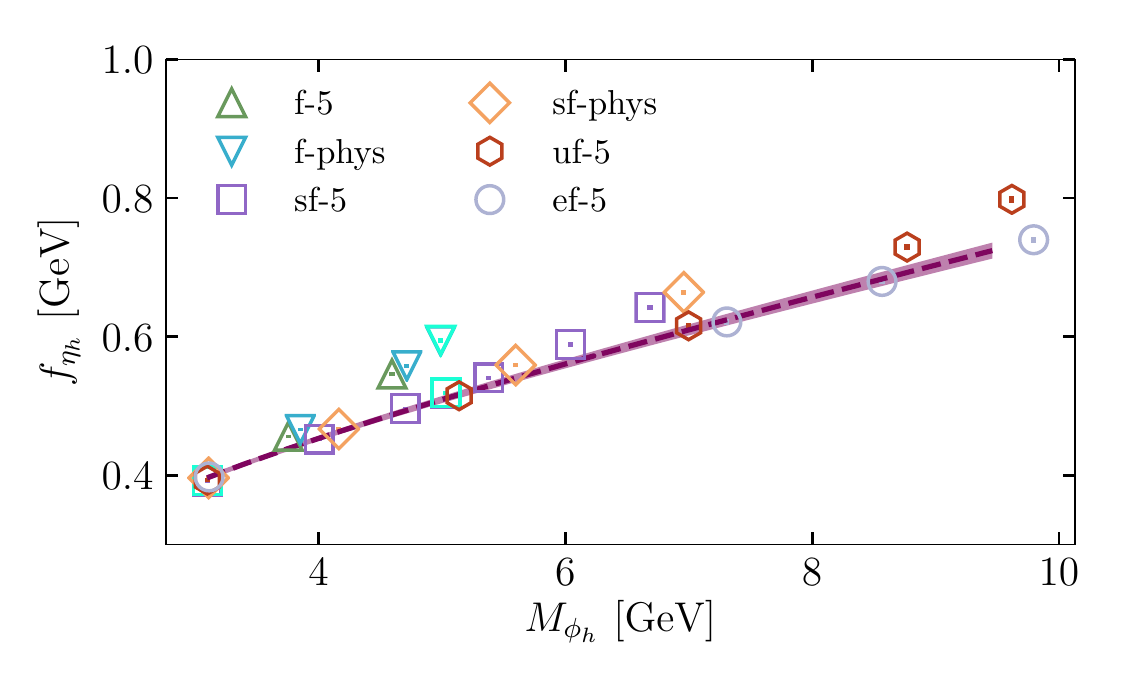}\\
\includegraphics[trim=0 5mm 0 0, clip, width=0.99\columnwidth]{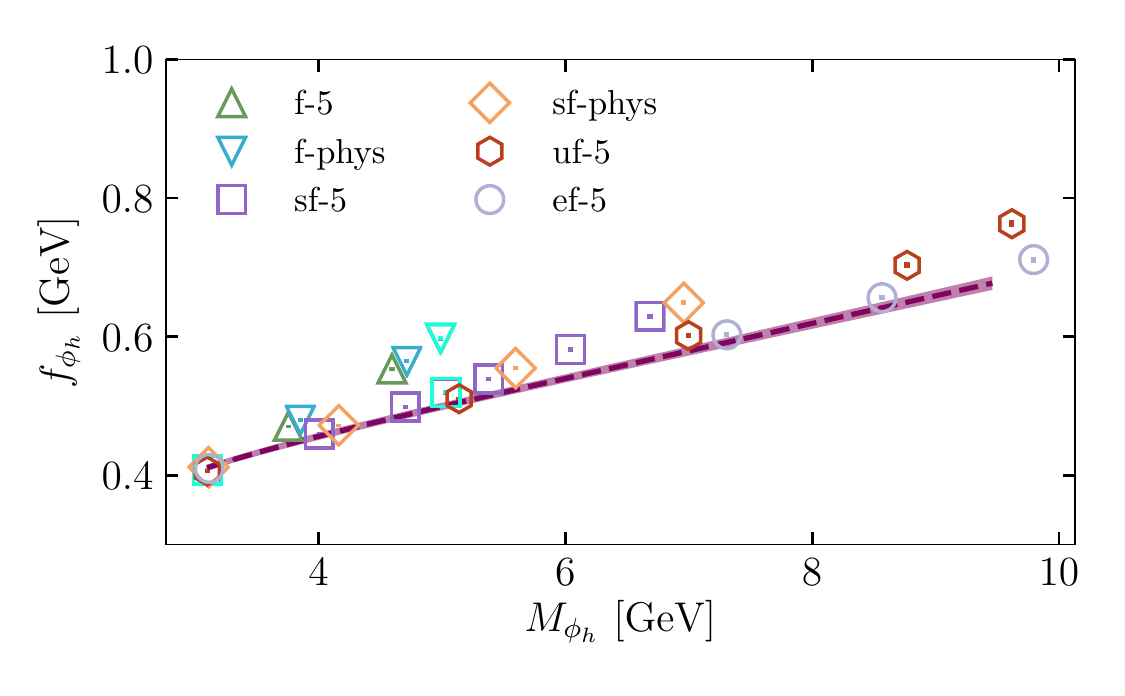}
\captionof{figure}{Bottomonium decay constants. These figures are from \cite{Hatton:2021dvg}.\label{fig:bottomonium_fP_fV}}
\endgroup

For bottomonium, we map the dependence of the pseudoscalar decay constant $f_{\eta_h}$ and the vector decay constant $f_{\phi_h}$
on the heavy quark mass, and extrapolate to the continuum and physical masses in the same way as for the hyperfine splitting.
This is illustrated in figure \ref{fig:bottomonium_fP_fV}, that shows lattice results from individual ensembles as well as
the extrapolation for both decay constants as a function of the vector meson mass $M_{\phi_h}$. The results at the
physical point are \cite{Hatton:2021dvg}
$f_{\Upsilon}=677.2(9.7)\mathrm{~MeV}$, $f_{\eta_b}=724(12)\mathrm{~MeV}$, and 
$f_{\Upsilon}/f_{\eta_b}=0.9454(99)$.
For charm the ratio $f_{J/\psi}/f_{\eta_c}$ is greater than 1, but for $b$ quarks this is
now shown to be $<1$.

As we briefly mentioned earlier, the partial decay width of a vector meson to a lepton pair
is directly related to the decay constant:
\begin{equation}
  \Gamma(\phi_h\to l^+l^-)=\frac{4\pi}{3}\alpha^2_{\textrm{QED}}Q^2\frac{f^2_{\phi_h}}{M_{\phi_h}},
\end{equation}  
where $Q$ is the electric charge of the quark. We can thus use our results for the vector
decay constants to calculate leptonic widths and compare with experiments, or vice versa.

\begingroup
\begin{center}
\includegraphics[trim={0 0.82cm 0 0},clip,width=0.76\columnwidth]{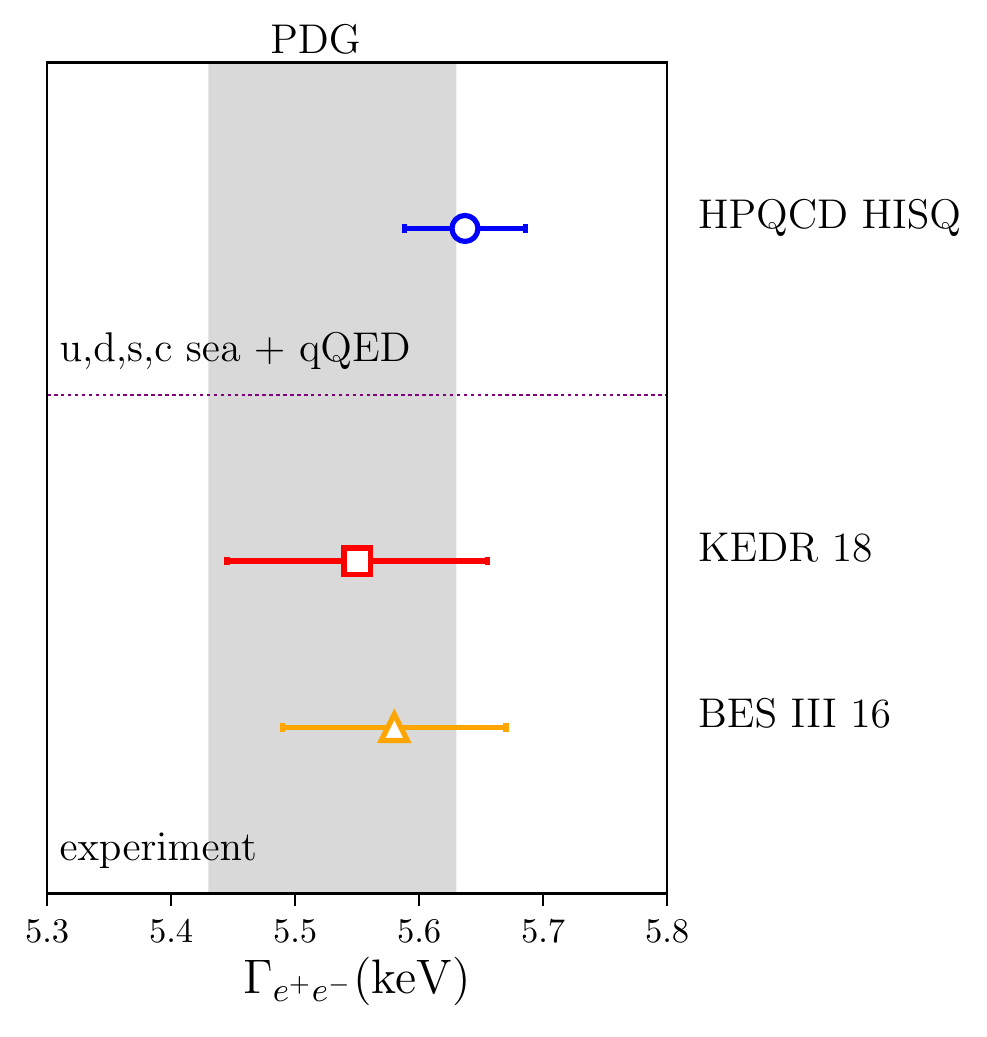}
\captionof{figure}{Leptonic width $\Gamma(J/\psi\to e^+e^-)$ [keV] (from \cite{Hatton:2020qhk}).
  \label{fig:charmonium_gamma_comparison}}
\end{center}
\endgroup

\begingroup
\begin{center}
\includegraphics[width=0.99\columnwidth]{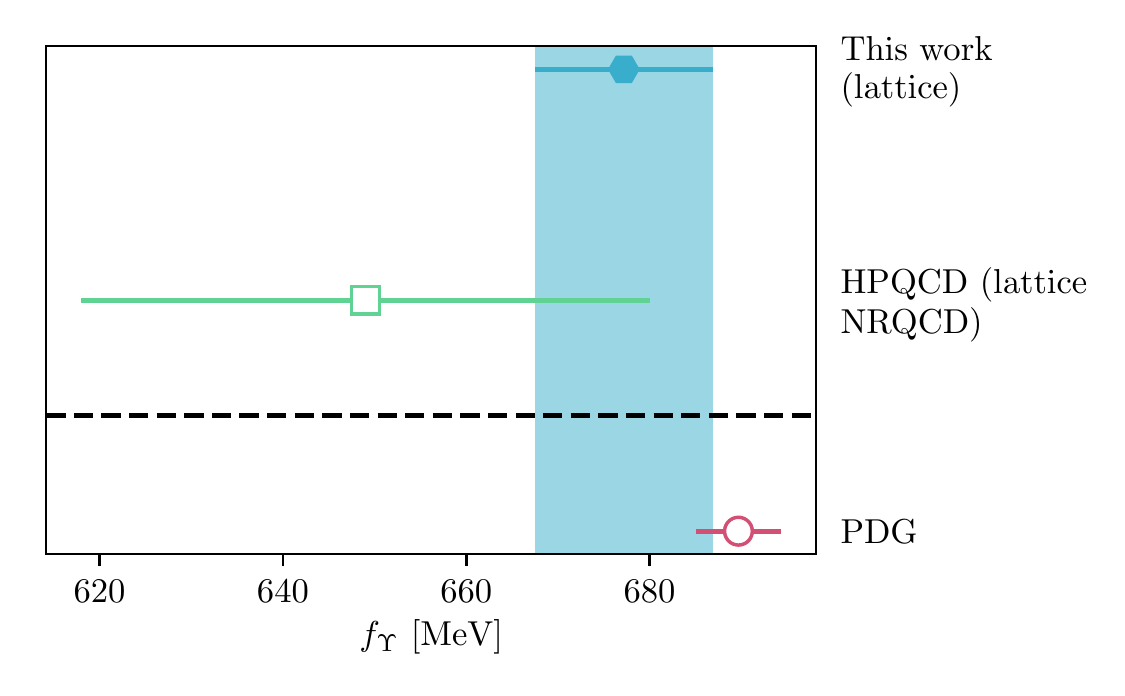}
\captionof{figure}{Bottomonium decay constant --- comparing lattice QCD (top result) with that inferred from experiment
  for $\Gamma(\Upsilon\to e^+e^-)$ (bottom result). This figure is from \cite{Hatton:2021dvg}.\label{fig:bottomonium_f_comparison}}
\end{center}
\endgroup

Our results are:
$\Gamma(J/\psi\to e^+e^-)=5.637(47)(13)\mathrm{~keV}$ and
$\Gamma(\Upsilon\to e^+e^-)=1.292(37)(3)\mathrm{~keV}$, and we show the comparison with
experiment in figures \ref{fig:charmonium_gamma_comparison} (charmonium) and
\ref{fig:bottomonium_f_comparison} (bottomonium). The agreement is seen to be good, and the
result from lattice for $\Gamma(J/\psi\to e^+e^-)$ is now more precise than the
experimental average from Particle Data Group. There is no experimental decay rate that can
be directly compared with the pseudoscalar decay constant.

We now turn to determining the $J/\psi$ tensor decay constant $f^T_{J/\psi}$. Recall that the
tensor decay constant is scale and scheme dependent, unlike the pseudoscalar and vector decay constants.
The calculation (published in \cite{Hatton:2020vzp}) can be summarised as follows:
\begin{enumerate}
\item Extract $\sqrt{2A_0^T/M_0^T}$ from tensor-tensor correlators.
  
\item
Calculate the renormalisation factor $Z_T^{\textrm{SMOM}}$.
Convert $f^T$ to the $\overline{MS}$ scheme at multiple scales
$\mu$ using the RI-SMOM scheme as an intermediate scheme on each ensemble.
  
\item
  Run all the $\overline{MS}$ tensor decay constants
  at a range of scales $\mu$ to a reference scale of
$2\textrm{ GeV}$ using a three-loop calculation of the
tensor current anomalous dimension. Here $\mu=2,3,4$~GeV.

\item
Fit all of the results for the $\overline{MS}$ decay constant at $2\textrm{ GeV}$
to a function that allows for discretisation effects
and non-perturbative condensate contamination coming from $Z_T^{\textrm{SMOM}}$.
\end{enumerate}

\begingroup
\begin{center}
\includegraphics[trim=5mm 5mm 2mm 2mm, clip, width=0.99\columnwidth]{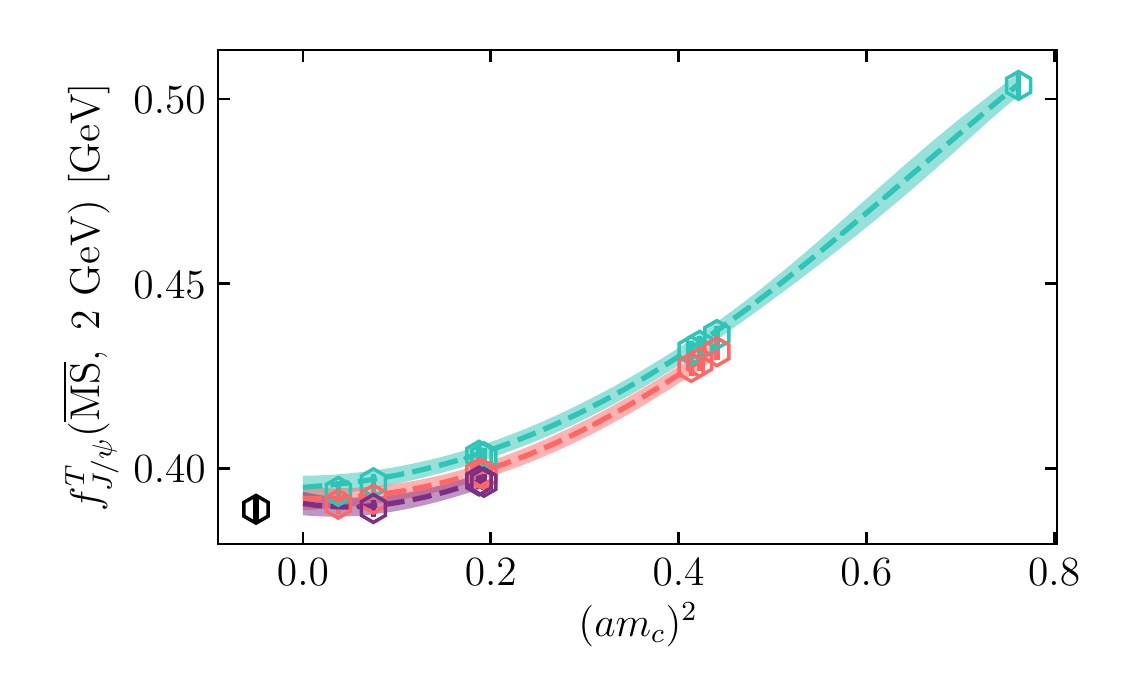}
\captionof{figure}{Tensor decay constant $f^T_{J/\psi}$. This figure is from \cite{Hatton:2020vzp}.\label{fig:Jpsi_tensor_f}}
\end{center}
\endgroup

The continuum extrapolation is illustrated in figure \ref{fig:Jpsi_tensor_f}. We
plot the tensor decay constant in the $\overline{MS}$ scheme at a scale of 2 GeV
using lattice tensor current renormalisation in the RI-SMOM scheme at
multiple $\mu$ values. These three values are shown as different coloured lines. The
blue line is 2 GeV, the orange, 3 GeV and the purple, 4 GeV.
The black hexagon is the physical result for $f^T_{J/\psi}(2\textrm{ GeV})$
obtained from the fit (with the condensate contamination removed).

In addition to the tensor decay constant $f^T_{J/\psi}(2\textrm{ GeV})$, we also determine
the ratio of the tensor and vector decay constants, $f^T_{J/\psi}/f^V_{J/\psi}$. The
extrapolation of the ratio to continuum is illustrated in figure \ref{fig:fT_fV_ratio}.
The colour coding for the lines and data points is the same as in figure
\ref{fig:Jpsi_tensor_f}.

\begingroup
\begin{center}
\includegraphics[trim=4mm 5mm 2mm 2mm, clip, width=0.9\columnwidth]{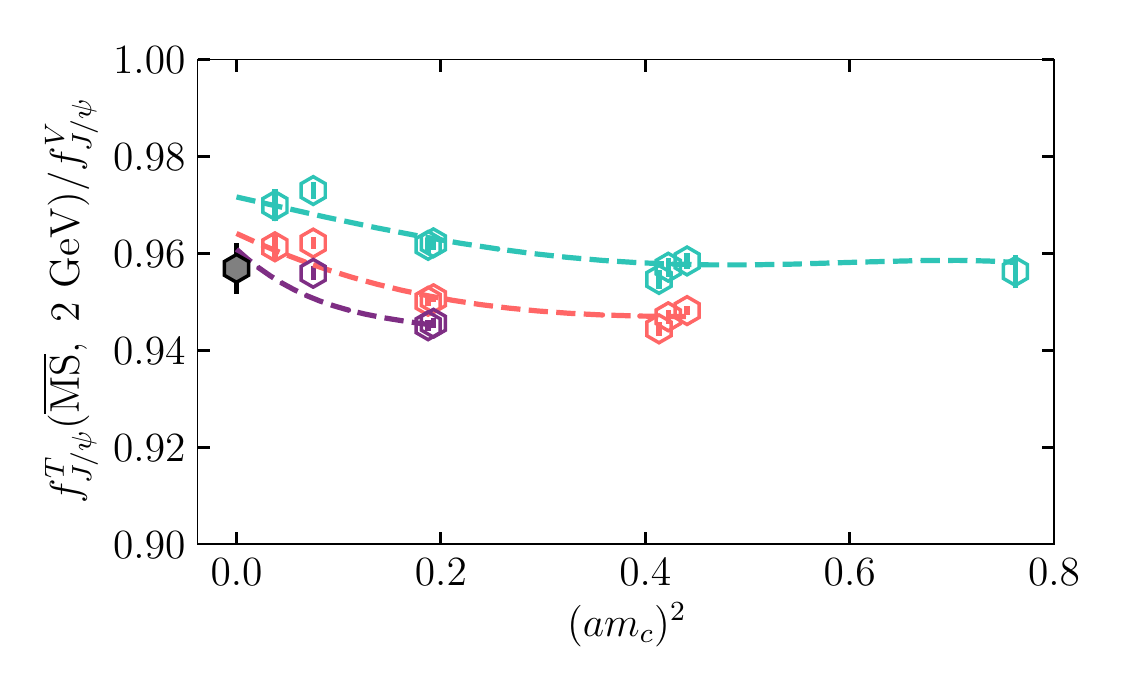}
\captionof{figure}{The ratio of tensor and vector decay constants. This figure is from \cite{Hatton:2020vzp}.\label{fig:fT_fV_ratio}}
\end{center}
\endgroup

Our (pure QCD) results for the $J/\psi$ tensor decay constant and its ratio with the vector decay constant
are \cite{Hatton:2020vzp}
$f^T_{J/\psi}(\overline{MS},2\,\mathrm{GeV})=0.3927(27)\mathrm{~GeV}$ and 
$f^T_{J/\psi}(\overline{MS},2\,\mathrm{GeV})/f^V_{J/\psi}=0.9569(52)$.
The ratio is compared to other lattice QCD and QCD sum rule calculations
\cite{Becirevic:2013bsa} in figure \ref{fig:fT_comparison}. Our result for the ratio is slightly (but
not significantly) lower than other results. The new determination of $f^T_{J/\psi}$ is much
more precise than the previous determinations. This is potentially useful for tests of BSM physics.

\begingroup
\begin{center}
\includegraphics[trim=6mm 3mm 4mm 2mm, clip, width=0.94\columnwidth]{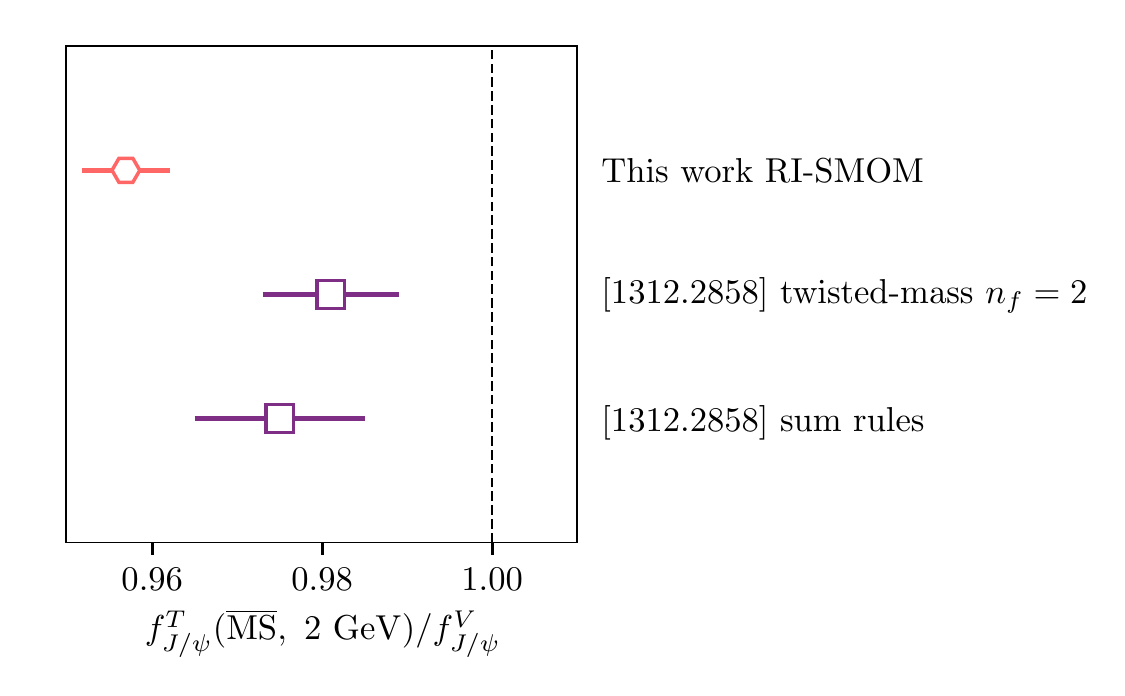}
\captionof{figure}{Comparison of the ratio of tensor and vector decay constants.. This figure is from \cite{Hatton:2020vzp}.\label{fig:fT_comparison}}
\end{center}
\endgroup

HPQCD's results show the high precision achievable now for the properties of ground-state
heavyonium mesons. In future this precision will be extended up the spectrum to excited states.

\section{Acknowledgements}

Computing was done on the Darwin supercomputer and on the Cambridge service for Data Driven Discovery (CSD3),
part of which is operated by the University of Cambridge Research Computing on behalf of the
DIRAC HPC Facility of the Science and Technology Facilities Council (STFC). The DIRAC component of CSD3
was funded by BEIS capital funding via STFC capital grants ST/P002307/1 and ST/R002452/1 and STFC
operations grant ST/R00689X/1. DiRAC is part of the national e-infrastructure. We are grateful to the
support staff for assistance. AL acknowledges support by the U.S. Department of Energy under grant number DE-SC0015655.

\end{multicols}
\medline
\begin{multicols}{2}

\end{multicols}
\end{document}